\documentclass[twocolumn,aps,prx,superscriptaddress]{revtex4-2}
\usepackage[T1]{fontenc}
\usepackage[latin9]{inputenc}
\usepackage{textcomp}
\usepackage{bbding}
\usepackage{amsmath}
\usepackage{amssymb}
\usepackage{graphicx}
\usepackage{esint}
\usepackage{xcolor}
\usepackage[pdftex,colorlinks,linkcolor=blue,citecolor=blue,urlcolor=blue,filecolor=blue,breaklinks=true]{hyperref}

\makeatletter
\@ifundefined{textcolor}{}
{%
 \definecolor{BLACK}{gray}{0}
 \definecolor{WHITE}{gray}{1}
 \definecolor{RED}{rgb}{1,0,0}
 \definecolor{GREEN}{rgb}{0,1,0}
 \definecolor{BLUE}{rgb}{0,0,1}
 \definecolor{CYAN}{cmyk}{1,0,0,0}
 \definecolor{MAGENTA}{cmyk}{0,1,0,0}
 \definecolor{YELLOW}{cmyk}{0,0,1,0}
}

\makeatother

\begin{document}
\title{Quantum optical neural networks using atom-cavity interactions to provide all-optical nonlinearity}
\author{Chuanzhou Zhu}
\email{chuanzhouzhu@arizona.edu}
\affiliation{Wyant College of Optical Sciences, University of Arizona, Tucson, Arizona, USA}
\author{Tianyu Wang}
\affiliation{Department of Electrical \& Computer Engineering, Boston University, Boston, Massachusetts, USA}
\author{Peter L. McMahon}
\affiliation{School of Applied and Engineering Physics, Cornell University, Ithaca, New York, USA}
\author{Daniel Soh}
\email{danielsoh@arizona.edu}
\affiliation{Wyant College of Optical Sciences, University of Arizona, Tucson, Arizona, USA}

\date{\today}

\begin{abstract}

Optical neural networks (ONNs) have been developed to enhance processing speed and energy efficiency in machine learning by leveraging optical devices for nonlinear activation and establishing connections among neurons. In this work, we propose a quantum optical neural network (QONN) that utilizes atom-cavity neurons with controllable photon absorption and emission. These quantum neurons are designed to replace the electronic components in ONNs, which typically introduce delays and substantial energy consumption during nonlinear activation. To evaluate the performance of the QONN, we apply it to the MNIST digit classification task, considering the effects of photon absorption duration, random atom-cavity detuning, and stochastic photon loss. Additionally, we introduce a convolutional QONN to facilitate a real-world satellite image classification (SAT-6) task. Due to its compact hardware and low power consumption, the QONN offers a promising solution for real-time satellite sensing, reducing communication bandwidth with ground stations and thereby enhancing data security.

\end{abstract}

\maketitle

\section{Introduction}

Recent advancements in artificial intelligence have been driven by the rapid growth in the number of internal parameters, such as the trillions of parameters in models like GPT-5. This expansion has led to a substantial increase in both energy consumption and the demand for faster processing speeds \cite{LeCun2015, thompson2022}.

Optical neural networks (ONNs) have been proposed as a potential solution to enhance processing speed and carrier frequency, benefiting from the broader optical bandwidth (typically in the terahertz range) compared to the electronic bandwidth (typically in the gigahertz range) \cite{Kazanskiy2022, Shen2017, Lin2018, Rios2019, Wetzstein2020, Xu2021, Feldmann2021, Zhou2021, Wang2022, Ashtiani2022, Sludds2022, Shastri2021, Hamerly2019, Nahmias2020, Bernstein2023, Varri2024, Lugnan2025, Choi2024}. ONNs have also achieved better energy efficiency than traditional electronic processors. In typical low-energy ONNs, millions of photons are necessary for the neuron pre-activation signal to ensure reliable performance \cite{Feldmann2021, Zhou2021, Sludds2022, Bernstein2023}. Recent advancements in ONN energy efficiency have reduced photon usage to hundreds of photons per activation \cite{Wang2022}. In the regime of ultra-low photon numbers, stochastic optical neural networks have been implemented to perform accurate machine learning inference, despite the challenges posed by low signal-to-noise ratios \cite{Ma2025}. The energy efficiency in these implementations is often measured by the number of photons involved in optical matrix-vector multiplications, which are performed through parallel vector-vector dot products enabled by techniques such as wavelength multiplexing \cite{Feldmann2021, Xu2021, Tait2019}, spatial multiplexing in photonic integrated circuits \cite{Shen2017, Stark2020, Bogaerts2020, Wu2021}, and spatial multiplexing in 3D free-space optical processors \cite{Miscuglio2020, Goodman1978, Psaltis1988, Dong2020, Chang2018, Matthes2019, Bueno2018, Spall2020}. 
ONNs have found applications in quantum information processing, particularly in quantum measurement tasks. Recently, an all-optical ONN based on electromagnetically induced transparency was developed for quantum state tomography. In this scheme, each nonlinear activation unit is engineered by utilizing the input laser intensity to modulate the transparency of laser-cooled rubidium atoms, resulting in a nonlinear relationship between the input and output optical intensities \cite{Zuo2022}.

However, two key sources of speed delay and energy consumption remain largely unaddressed in ONNs: the nonlinear activation following a multiplication operation and the photon emission required before the next multiplication stage. Both processes rely on macroscopic electronic devices, such as single-photon detectors and photon emitters \cite{Wang2022, Ma2025}, which lead to the electronic delay in processing speed and contribute significantly to overall energy consumption at the macroscopic level. 

Quantum neural networks (QNNs) \cite{Liao2024, Wen2024, Bischof2025, Hirai2024, Thompson2025, Garcia2025, Chen2024, Cong2019, Moreira2023, Zhang2024, Pan2023, Zhu2022, Beer2020, Killoran2019} have been proposed as quantum generalizations of classical neural networks, implemented on small-scale quantum platforms such as superconducting quantum processors \cite{Moreira2023, Pan2023}. These architectures might provide benefits due to uniquely quantum phenomena such as entanglement. A QNN comprises a set of parameterized quantum logical gates that perform nonlinear activation in each neuron and produce quantum entanglement across multiple neurons. To achieve the desired outputs from QNNs, the gate parameters are typically trained using classical \cite{McClean2016} and quantum \cite{Liao2024} optimizers. However, quantum decoherence increases the complexity of designing reliable quantum gates \cite{Sar2012, Xue2022}, particularly for two-qubit gates \cite{Miifmmode2020, Huang2018}, which has spurred efforts to develop fault-tolerant quantum gates through topological protection \cite{Kitaev2006, Choi2022} and error correction \cite{Roffe2019}. Additionally, the need for fine-tuning internal gate parameters further complicates the fabrication of robust QNNs.

Quantum optical neural networks (QONNs) \cite{Steinbrecher2019, Ewaniuk2023, Roncallo2025, Killoran2019} have emerged as novel platforms for implementing linear transformations and nonlinear activation functions in neural networks through quantum optical systems. 
Compared with ONNs, QONNs employ small-scale quantum optical components in place of the conventional macroscopic electronic devices used for nonlinear activation, thereby further reducing latency and power consumption. In contrast to QNNs, QONNs use optical signals to transmit information between layers, which helps mitigate information loss caused by quantum decoherence and dissipation.
Although these QONNs do not rely directly on qubit gates as in QNNs, they still employ gate components such as continuous-variable gates \cite{Killoran2019} and optical unitaries \cite{Steinbrecher2019, Ewaniuk2023}, along with internal gate parameters that must be precisely adjusted. Actually, even before machine learning and artificial intelligence entered the public consciousness, quantum-optics-based neural networks had already been proposed to simulate information processing in neurobiological experiments. Early examples include a network composed of two neural states exhibiting transitions analogous to photon-emission models in quantum optics \cite{Lewenstein1991, Lewenstein1992}, a quantum network hardware architecture constructed from a spatial array of quantum dot molecules \cite{Behrman1999}, and an optical neural network built with quantum well devices employing two arrays of integrated asymmetric Fabry-Perot modulators \cite{Jennings1994}. More recently, a QONN has been proposed by mapping the features of modern machine-learning neural networks onto the few-photon quantum optical domain \cite{Steinbrecher2019}. In this framework, the nonlinear activation is realized through a phase quadratic generated via single-mode Kerr interactions, while the matrix multiplication is implemented using a linear optical unitary achieved through arrays of beam splitters and programmable phase shifts \cite{Reck1994}. Subsequent studies have investigated the effects of photon propagation loss, weak nonlinearities, and network size on QONN performance \cite{Ewaniuk2023}. Furthermore, it has been proposed that nonlinear activation in QONNs can also be produced using the Hong-Ou-Mandel effect \cite{Roncallo2025}. However, in QONNs that rely on the Kerr effect and optical unitaries, the performance is constrained by the weak Kerr nonlinearity in the few-photon regime \cite{Matsuda2009}. Moreover, tuning the trainable parameters in the optical unitary does not enable modulation of the average photon amplitude.

We present a QONN that employs atom-cavity neurons to replace the electronic photon detectors and emitters in ONNs, while avoiding the use of quantum gates or the adjustment of internal gate parameters as seen in QNNs \cite{Liao2024, Wen2024, Bischof2025, Hirai2024, Thompson2025, Garcia2025, Chen2024, Cong2019, Moreira2023, Zhang2024, Pan2023, Zhu2022, Beer2020, Killoran2019} and previous proposals for QONNs \cite{Steinbrecher2019, Ewaniuk2023, Roncallo2025, Killoran2019}. Our QONN harnesses cavity arrays \cite{Zhu2013, Dong2015, Zhu2016} and optical matrix-vector multipliers (MVMs) to transmit single-photon amplitudes, which serve as information carriers across multiple layers of the network. Each neuron in the cavity array performs nonlinear activation on the optical signal and is constructed with an atom placed inside two cavities, which can be switched on and off to absorb and emit photons. 
Our results show that the desired nonlinearity in the activation function can be achieved by tuning the photon absorption duration.
The optical MVM is implemented via a controllable spatial light modulator (SLM) that performs fully connected transformations or convolution operations. 
Experimental results have demonstrated that flexible parameter adjustment in optical MVMs enables high inference accuracy in ONNs \cite{Wang2022, Ma2025}.

To demonstrate the capability of QONN in image recognition, we adopt MNIST handwritten digit classification \cite{Deng2012} as a benchmark task to identify the optimal photon absorption duration for each cavity array that performs as a hidden layer. This is followed by simulations under realistic conditions, incorporating random atom-cavity detuning in each cavity neuron and stochastic single-photon loss during photon transmission. To highlight the real-world applicability of QONN in remote sensing, we test the benchmarked system on a satellite image classification task, namely the DeepSat (SAT-6) airborne image classification \cite{Basu2015}. To further optimize the system, we propose a convolutional QONN to reduce the number of controllable SLM pixels without compromising accuracy or requiring additional cavity neurons.

To derive an analytical formalism for the atom-cavity-based all-optical nonlinear activation function, we employ a semiclassical mean-field treatment of the photon field, where the photon-photon entanglement in optical MVMs is not included. Our proposed nonlinear activation mechanism for the photon amplitude is based on the Rabi dynamics of a switchable cavity-QED system consisting of two coupled cavities, with the degree of nonlinearity conveniently tunable through the photon absorption duration. Although performance enhancements associated with quantum entanglement have been reported in other quantum artificial intelligence frameworks, such as quantum reservoir computing \cite{kornjaca2024, Zhu2025, Zhu2024}, there is currently no clear evidence of a strong correlation between entanglement and performance in QONN architectures, including QONN models that go beyond the mean-field approximation by incorporating photon Fock states in continuous-variable gates and optical unitary transformations \cite{Steinbrecher2019, Ewaniuk2023, Roncallo2025, Killoran2019}.

\begin{figure*}
\includegraphics[width=1.0\linewidth]{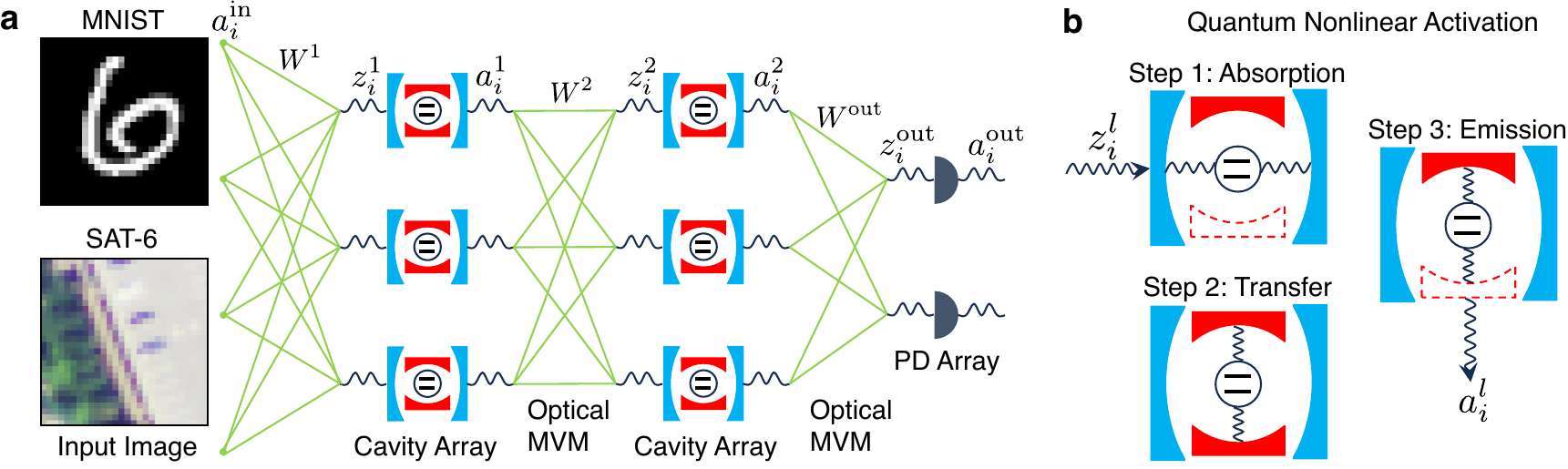}
\caption{Quantum optical neural network for MNIST digit classification and DeepSat (SAT-6) airborne image classification tasks. 
{\bf a} Schematic of the quantum optical neural network, consisting of optical matrix-vector multipliers (MVMs) and cavity arrays, where $z_{i}^{l}$ represents the incident photon amplitude entering each cavity neuron, and $a_{i}^{l}$ is the photon amplitude emitted by the atom inside the cavity neuron. The activation values $a_{i}^{l}$ from the cavity array of each layer are passed to the next layer through the optical MVM, with $z_{i}^{l+1}$ being the weighted sum of $a_{i}^{l}$. The optical MVM is implemented using a spatial light modulator (SLM). The weights $W_{ij}^{l}$, which are parameters optimized via backpropagation, are adjusted by controlling the transmission of each SLM pixel. The input images are encoded in $a_{i}^{\rm{in}}$. The final layer uses a photodetector (PD) array to obtain the output $a_{i}^{\rm{out}}$, which is used to compute the cost function. 
{\bf b} Schematic of a cavity neuron that performs the quantum optical nonlinear activation as described by Eq.~(\ref{eq:activation}). The neuron consists of a low-Q cavity with weaker atom-cavity coupling (in blue) and a high-Q cavity with stronger atom-cavity coupling (in red). In step 1, the incident photon is absorbed by the two-level atom, and the excitation is stored in the low-Q cavity while the high-Q cavity is turned off. In step 2, the high-Q cavity is turned on, and the excitation is transferred from the low-Q cavity to the high-Q cavity. In step 3, the high-Q cavity is re-opened, and the photon is emitted by the atom through a complete energy conversion from atomic excitation to photon emission.}
\label{Fig1}
\end{figure*}

\section{Setup}

The schematic of the QONN for image recognition is illustrated in Fig.~\ref{Fig1}a. It consists of an input layer, two fully connected hidden layers, and an output layer, where the layers are indexed by $l$ and the neurons in each layer by $i$. The framework can be extended to include additional hidden layers to further enhance performance. Each pixel in an input image is encoded as a photon amplitude, denoted by $a_{i}^{{\rm in}}$. The optical matrix-vector multiplier (MVM), implemented via a spatial light modulator (SLM), linearly connects the activation function $a_{i}^{l-1}$ of the $(l-1)$-th layer to the incident photon amplitude $z_{i}^{l}$ of the $l$-th layer. The controllable transmission of each SLM pixel realizes the trainable parameter $W_{ij}^{l}$ in the network. The cavity array, composed of decoupled cavity neurons, performs quantum optical activation by establishing a nonlinear relation between each incident photon amplitude $z_{i}^{l}$ and the emitted photon amplitude $a_{i}^{l}$. The photodetector (PD) array in the final layer retrieves the output photon amplitude $a_{i}^{{\rm out}}$, which corresponds to the inference result and is utilized to calculate the cost function.

Quantum nonlinear activation is realized through the cavity neuron illustrated in Fig.~\ref{Fig1}b. Each neuron comprises a two-level atom surrounded by two optical cavities: one with a lower Q factor and weaker atom-cavity coupling strength (blue), and another with a higher Q factor and stronger atom-cavity coupling (red). The high-Q cavity can be switched on or off by suppressing or enhancing photon loss using a phase shifter, which controls the destructive interference between the leftward and rightward traveling waves of the cavity mirror system \cite{Soh2021, Taylor2022}. The nonlinear activation process involves three steps: photon absorption, excitation transfer, and spontaneous emission.
 
\textit{Step 1}: The high-Q cavity is turned off, and the low-Q cavity facilitates photon absorption by the two-level atom. The incident photon acts as an external drive for the atom. For the $i$-th cavity neuron in the $l$-th layer, the absorption process is governed by the Hamiltonian
\begin{equation}
H_{i}^{l}=\delta_{i}^{l}\sigma_{z}+g\left(\sigma_{+}z_{i}^{l}+\sigma_{-}z_{i}^{l}\right),\label{eq:Hli}
\end{equation}
where $z_{i}^{l}$ is the amplitude of the incident photon field under the mean-field approximation, $\delta_{i}^{l}$ denotes the atom-cavity detuning, $g$ represents the atom-photon electric dipole coupling strength, and $\sigma_{z}$ and $\sigma_{\pm}$ are Pauli matrices. Each atom is initialized in the ground state, and the atomic excitation follows a Rabi oscillation described by
\begin{equation}
\left\langle \sigma_{z}\right\rangle _{i}^{l}=-\left(\frac{\delta_{i}^{l}}{\Omega_{i}^{l}}\right)^{2}-\left(\frac{gz_{i}^{l}}{\Omega_{i}^{l}}\right)^{2}\cos\left(2\pi t_{l}\Omega_{i}^{l}\right),\label{eq:sigmazli}
\end{equation}
where the oscillation frequency depends on $z_{i}^{l}$ as
\begin{equation}
\Omega_{i}^{l}\left(z_{i}^{l}\right)\equiv\sqrt{\left(gz_{i}^{l}\right)^{2}+\left(\delta_{i}^{l}\right)^{2}},\label{eq:Omegali}
\end{equation}
and $t_{l}$ is the absorption time, which is assumed identical for all neurons in the same layer. 

\textit{Step 2}: The high-Q cavity is turned on. Since the atom tends to more rapidly interact with the high-Q cavity through its photon absorption and emission within the two-cavity neuron, the excitation is transferred from the low-Q cavity to the high-Q one. 

\textit{Step 3}: Once all cavity neurons in a layer are excited, all high-Q cavities in that layer are simultaneously opened. Through complete energy conversion from atomic excitation to photons, the intensity of the spontaneously emitted photon field is given by
\begin{equation}
\left|a_{i}^{l}\right|^{2}=\frac{1}{2}\left(\left\langle \sigma_{z}\right\rangle _{i}^{l}+1\right),\label{eq:alisquare}
\end{equation}
where $a_{i}^{l}$ represents the emitted photon amplitude. Accounting for a random phase $\phi_{i}^{l}$ induced by spontaneous emission, the emitted photon amplitude is expressed as
\begin{equation}
a_{i}^{l}=\left|a_{i}^{l}\right|e^{i\phi_{i}^{l}},\label{eq:ali1}
\end{equation}
where $\left|a_{i}^{l}\right|$ follows Eq.~(\ref{eq:alisquare}) and $\left\langle \sigma_{z}\right\rangle _{i}^{l}$ is defined in Eq.~(\ref{eq:sigmazli}).

The activation values $a_{i}^{l}$ are linearly propagated to the next layer through optical MVM, implemented by displaying these values on the SLM, as shown in Fig.~\ref{Fig1}a. The SLM consists of a pixel array with individually controllable transmission for each pixel, as demonstrated in Refs.~\cite{Wang2022, Ma2025}. Element-wise multiplication is achieved through amplitude modulation, resulting in the linear combination
\begin{equation}
z_{i}^{l}=\underset{j}{\sum}W_{ij}^{l}a_{j}^{l-1},\label{eq:zli}
\end{equation}
where each weight $W_{ij}^{l}$ is a complex number controlled by a single SLM pixel with both amplitude and phase modulations. In an ONN experiment \cite{Ma2025}, the magnitude and phase of $W_{ij}^{l}$ correspond to the optical attenuation and phase shift applied to the incident field, respectively. Our model restricts the weight within $W_{ij}^{l}\in[-1,1]$, which is equivalent to setting its phase to $0$ or $\pi$.
These weights serve as trainable parameters of the neural network, which are optimized during the training process described in Appendix \ref{sec:Backpropagation}.

To eliminate the randomness of the phase $\phi_{i}^{l}$ in the activation function (\ref{eq:ali1}), a weak auxiliary external laser need to be applied to induce stimulated emission for phase locking \cite{Bazarov1987, Likhanskii_1991}. The locked phase $\phi_{i}^{l}$ can then be compensated by the complex phase of the transmission rate $W_{ij}^{l}$, allowing $a_{i}^{l}$, $z_{i}^{l}$, and $W_{ij}^{l}$ to be treated as real-valued. Consequently, the auxiliary phase-locking process yields the nonlinear quantum activation function
\begin{equation}
a_{i}^{l}\left(z_{i}^{l}\right)=\frac{g\left|z_{i}^{l}\right|}{\Omega_{i}^{l}\left(z_{i}^{l}\right)}\left|\sin\left[\pi t_{l}\Omega_{i}^{l}\left(z_{i}^{l}\right)\right]\right|,\label{eq:activation}
\end{equation}
where $\Omega_{i}^{l}\left(z_{i}^{l}\right)$ is a nonlinear function of $z_{i}^{l}$ defined by Eq.~(\ref{eq:Omegali}). The pre-activation and activation values span $z_{i}^{l}\in\left(-\infty,+\infty\right)$ and $a_{i}^{l}\in\left[0,1\right]$, respectively.
The degree of nonlinearity of this cavity-assisted activation is controllable through tuning the photon absorption duration $t_{l}$ in each layer, enabling operation beyond the weak-nonlinearity regime considered in \cite{Steinbrecher2019, Ewaniuk2023}. The effects of photon absorption duration are discussed in Section \ref{Sec_Photon_Absorption_Durations}.

The design of our two-cavity neuron supports temporal multiplexing for optically decomposing the matrix-vector multiplication in Eq.~(\ref{eq:zli}) into a batch of vector-vector dot products, which are executed in time sequence. For instance, the input photon amplitudes $a_{i}^{{\rm in}}$ are transmitted through the SLM pixels and focused onto the first neuron in the first hidden layer, leading to the vector-vector dot product $z_{1}^{1}=\sum_{j}W_{1j}^{1}a_{j}^{{\rm in}}$. 
The resulting $z_{1}^{1}$ corresponds to the effective classical field amplitude in the Hamiltonian $H_{1}^{1}$ in \textit{Step 1}, as defined in Eq.~(\ref{eq:Hli}). Since the summation includes contributions from a large number of neurons in the preceding layer (e.g., hundreds of neurons in the input layer, as shown in Fig.~\ref{Fig1}(a)), the quantum fluctuations associated with the individual single-photon amplitudes $a_{j}^{{\rm in}}$ are averaged out. Consequently, $z_{1}^{1}$ can be regarded as an effective classical optical field amplitude.
The excitation is then held during \textit{Step 2} by maintaining the high-Q cavity (red) in the ``on'' state, preserving the excitation inside the high-Q cavity without photon emission. This procedure sequentially excites all neurons in the first hidden layer, corresponding to $z_{i}^{1}=\sum_{j}W_{ij}^{1}a_{j}^{{\rm in}}$ for each of $i$-th neurons in the layer 1. After the entire first hidden layer is excited, all neurons in this layer simultaneously proceed to \textit{Step 3}, during which the high-Q cavities are opened and all emitted photons are transmitted to the first neuron in the second hidden layer, producing $z_{1}^{2}=\sum_{j}W_{1j}^{2}a_{j}^{1}$. This temporal multiplexing continues throughout the network and achieves a fan-out mechanism, in which one neuron's single emitted photon reaches many next-layer neurons.

The output layer employs an array of phase-sensitive photodetectors to measure the output amplitudes $z_{i}^{{\rm out}}$. Such phase-sensitive measurements are typically achieved through interferometric techniques, where the incoming optical signal is mixed with a coherent reference beam to retrieve the amplitude and phase information of the optical field \cite{Choi2007}. The measured logits, $z_{i}^{{\rm out}}$, are then processed by the LogSoftmax function to obtain the normalized outputs, represented as $a_{i}^{{\rm out}}$. The deviation between the predicted outputs and the ground truth is quantified using the cross-entropy loss, which serves as the cost function for network optimization.
The number of photodetectors corresponds to the number of categories in the specific classification task. In the MNIST digit recognition task, the target output is defined as $a_{i=digit}^{{\rm out}}=1$ and $a_{i\ne digit}^{{\rm out}}=0$, where $digit$ denotes the input handwritten numeral. The neural network is trained using backpropagation with the stochastic gradient descent (SGD) optimizer. Further details regarding the training and testing procedures are provided in Appendix \ref{sec:Backpropagation}.

Our setup eliminates the need for thousands of single-photon detectors and emitters typically required in the hidden layers in previous ONN experiments \cite{Wang2022, Ma2025}. The energy consumption and speed of each single-photon detector are roughly estimated as $1\rm{W}$ and $100\rm{MHz}$, respectively. Our setup requires only a small number of photon detectors at the output layer (e.g. ten detectors for the MNIST task), thereby reducing the overall energy consumption and potentially improving the processing speed. In the proposed model, the direct implementation of a SLM panel for each layer can cost several $\rm{Watt}$, but the linear transformation can also be achieved using a Mach-Zehnder interferometer network, which reduces the power consumption to the photon energy range ($10^{-19}\rm{J}$). The energy costs of other components, including the cavity arrays and the weak auxiliary lasers for phase locking, are also in the photon energy range. The processing speed of the cavity neuron is determined by the Rabi frequency, which typically falls within the tens of $\rm{MHz}$ range in conventional atom-photon coupling experiments \cite{Johnson2008, Choi2025}. In contrast, all-optical setups involving photonic qubits can achieve Rabi frequencies in the $\rm{THz}$ range \cite{Lu2024, Yumoto2018, Leitenstorfer2014}, offering the potential to substantially accelerate processing speeds. The all-optical implementation of our cavity quantum electrodynamics neurons can be realized using quantum dots embedded in a photonic crystal \cite{Stockklauser2017, Englund2007, Winger2008}. The typical speed of a SLM is several $\rm{kHz}$ \cite{Pivnenko2021}, while the replacement with a Mach-Zehnder interferometer network can potentially boost the speed to $100\rm{GHz}$ \cite{Ataei2024}.

\begin{figure*}
\includegraphics[width=1.0\linewidth]{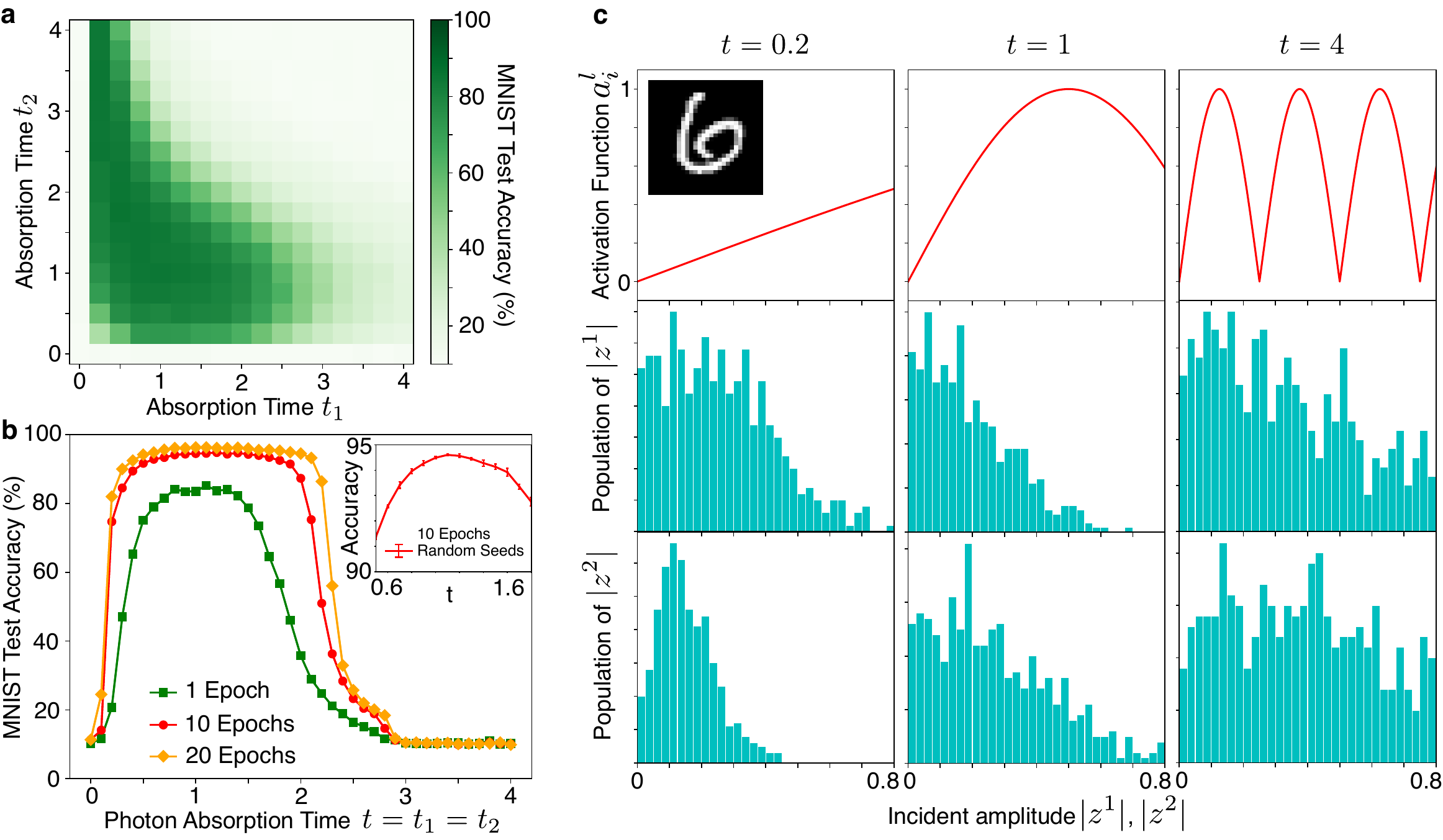}
\caption{
Performance of the quantum optical neural network with various photon absorption times for the MNIST task. 
{\bf a} Test accuracy plotted against the photon absorption times in the first and second hidden layers, $t_{1}$ and $t_{2}$. 
{\bf b} Test accuracy as a function of the absorption time $t$, with $t=t_{1}=t_{2}$, after 1 (green), 10 (red), and 20 (orange) training epochs. A training epoch is defined as a complete pass of the entire MNIST training dataset through the learning process. 
The inset in {\bf b} shows the optimal accuracy achieved by 10 training epochs, with the error bars representing the standard deviations across 5 independent runs with different random seeds.
{\bf c} Nonlinear activation function $a_{i}^{l}$ (identical for all neurons in the two layers since $t_{1}=t_{2}$), and the occurrence population distributions of the incident photon amplitudes $|z_{i}^{1}|$ and $|z_{i}^{2}|$, which serve as the arguments for the activation functions, The results shown in panel {\bf c} correspond to three absorption times, $t=0.2$, $t=1$, and $t=4$, based on $10$ training epochs and the input image shown in the upper-left corner.
The detuning is $\delta_{i}^{l}=0$, and the coupling strength is $g=1$ for every cavity neuron. The number of neurons is set to $N_{1} = N_{2} = 512$ in the two hidden layers.}
\label{Fig2}
\end{figure*}

\section{Performance Evaluation under Realistic Conditions}

\subsection{Photon Absorption Durations}
\label{Sec_Photon_Absorption_Durations}

According to the formalism of the activation function in Eq.~(\ref{eq:activation}), the photon absorption time $t_{l}$ in \textit{Step 1} of the quantum activation determines the degree of nonlinearity and monotonicity between the pre-activation $z_{i}^{l}$ and the activation value $a_{i}^{l}$. When the detuning $\delta_{i}^{l}=0$, the activation function simplifies to
\begin{equation}
a_{i}^{l}\left(z_{i}^{l}\right)=\left|\sin\left(\pi t_{l}g\left|z_{i}^{l}\right|\right)\right|.\label{eq:activation0}
\end{equation}
Based on the setup with two hidden layers, Fig.~\ref{Fig2}a illustrates the test accuracy for MNIST digit classification as a function of the photon absorption times in the first and second hidden layers, denoted by $t_{1}$ and $t_{2}$, respectively. When $t_{1}=t_{2}=t$, Fig.~\ref{Fig2}b presents the improvement of test accuracy as the number of training epochs increases (i.e., repeated iterations over the MNIST dataset). The performance tends to saturate after $10$ epochs. The dependence of accuracy on $t$ arises from the trade-off between the nonlinearity and monotonicity of the activation function. As shown in Fig.~\ref{Fig2}c, for a short absorption duration ($t=0.2$), most pre-activation values $z_{i}^{l}$ fall within the nearly linear regime of the activation function, rendering the activation insufficiently nonlinear to achieve high accuracy. Conversely, for a long absorption duration ($t=4$), the rapid oscillations of the activation function disrupt the monotonic correspondence between $z_{i}^{l}$ and $a_{i}^{l}$, such that different incident amplitudes produce identical photon emissions. This loss of one-to-one mapping leads to information degradation and reduced accuracy. The optimal performance occurs around an intermediate absorption time ($t=1$), where the nonlinearity and monotonicity are effectively balanced.

The photon absorption durations in the first and second hidden layers, $t_{1}$ and $t_{2}$, play distinct roles in determining accuracy. The asymmetric colormap in Fig.~\ref{Fig2}a indicates that optimal performance requires tighter control over $t_{1}$, whereas a broader range of $t_{2}$ values yields acceptable accuracies. This asymmetry can be understood from the parameter training process: once the activation functions in both layers are fixed by a given combination of $t_{1}$ and $t_{2}$, the training process can only adjust the weights $W_{ij}^{1}$ to optimize the distribution of neuron populations in the $z_{i}^{1}$ domain, thereby balancing the benefits of nonlinearity and monotonicity in the first layer. In contrast, for the second layer, the training process can further tune both $W_{ij}^{1}$ and $W_{ij}^{2}$ to reach the optimal population distribution of $z_{i}^{2}$ values. This interpretation is consistent with the distinct distributions of $z_{i}^{1}$ and $z_{i}^{2}$ illustrated in Fig.~\ref{Fig2}c. 

\begin{figure}
\includegraphics[width=1.0\linewidth]{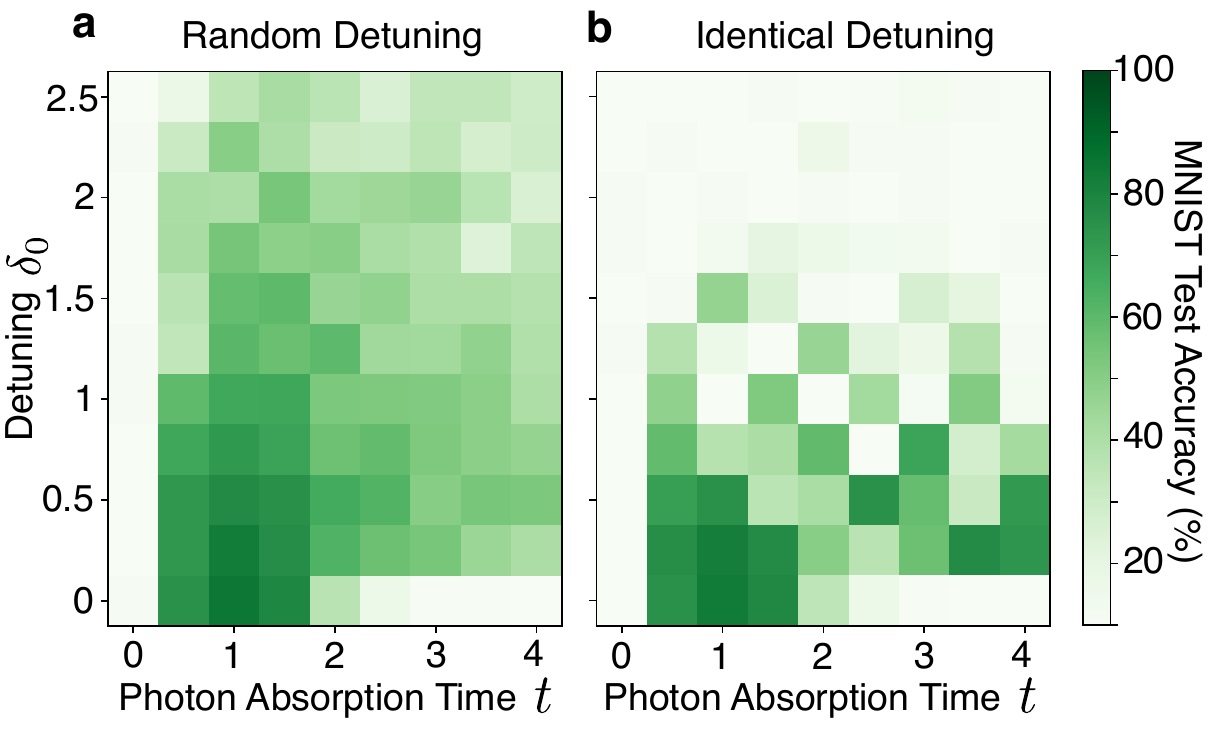}
\caption{Performance of the quantum optical neural network with finite detuning for each cavity neuron on the MNIST task. The photon absorption times $t$ are identical in the first and second hidden layers. 
{\bf a} Each cavity neuron is simulated with a random detuning $\delta_{i}^{l}$, where the probability is uniformly distributed within the range $[-2\delta_{0}, 2\delta_{0}]$, resulting in an average detuning magnitude of $\delta_{0}$.
{\bf b} All cavity neurons are identical and simulated with a fixed detuning $\delta_{i}^{l} = \delta_{0}$.
The coupling strength is $g=1$, and the number of neurons is $N_{1} = N_{2} = 512$.}
\label{Fig3}
\end{figure}

\subsection{Random Detuning}

In all-optical implementations using quantum dots embedded in photonic crystals to realize artificial two-level atoms inside optical cavities, the atom-photon detuning in each cavity is typically difficult to control and is often modeled as a random variable \cite{Stockklauser2017, Englund2007, Winger2008, Ledentsov1996}. Figure~\ref{Fig3}a presents the influence of such random detuning on the MNIST test accuracy, where the activation function of each neuron is simulated using Eq.~(\ref{eq:activation}) with a finite detuning $\delta_{i}^{l}$. The detuning $\delta_{i}^{l}$ for each neuron is independently drawn from a uniform probability distribution over the range between $-2\delta_{0}$ and $2\delta_{0}$. The QONN exhibits robust performance against a wide range of random detunings. It is worth noting that the values of random detuning shown in Fig.~\ref{Fig3}a (up to $\delta_{0}=2.5$) represent substantial magnitudes, given that the coupling strength is set to $g=1$ in the simulation and the typical magnitude of $\left|z_{i}^{l}\right|$ is on the order of $1$ as evidenced in Fig.~\ref{Fig2}c.

For comparison, Fig.~\ref{Fig3}b presents the case where all cavity neurons are identical and share the same detuning $\delta_{i}^{l}=\delta_{0}$. In this scenario, the network achieves better accuracy only for a few specific combinations of $\delta_{0}$ and $t$, corresponding to occasional optimal balances between nonlinearity and monotonicity in the activation function. In contrast, random detuning relaxes the requirements for precise control over both absorption time and detuning. The generally higher accuracy achieved under random detuning highlights the advantage of introducing variability among cavity neurons, enabling them to develop non-identical response characteristics and thereby enhancing the collective computational capability of the network.

\begin{figure}
\includegraphics[width=1.0\linewidth]{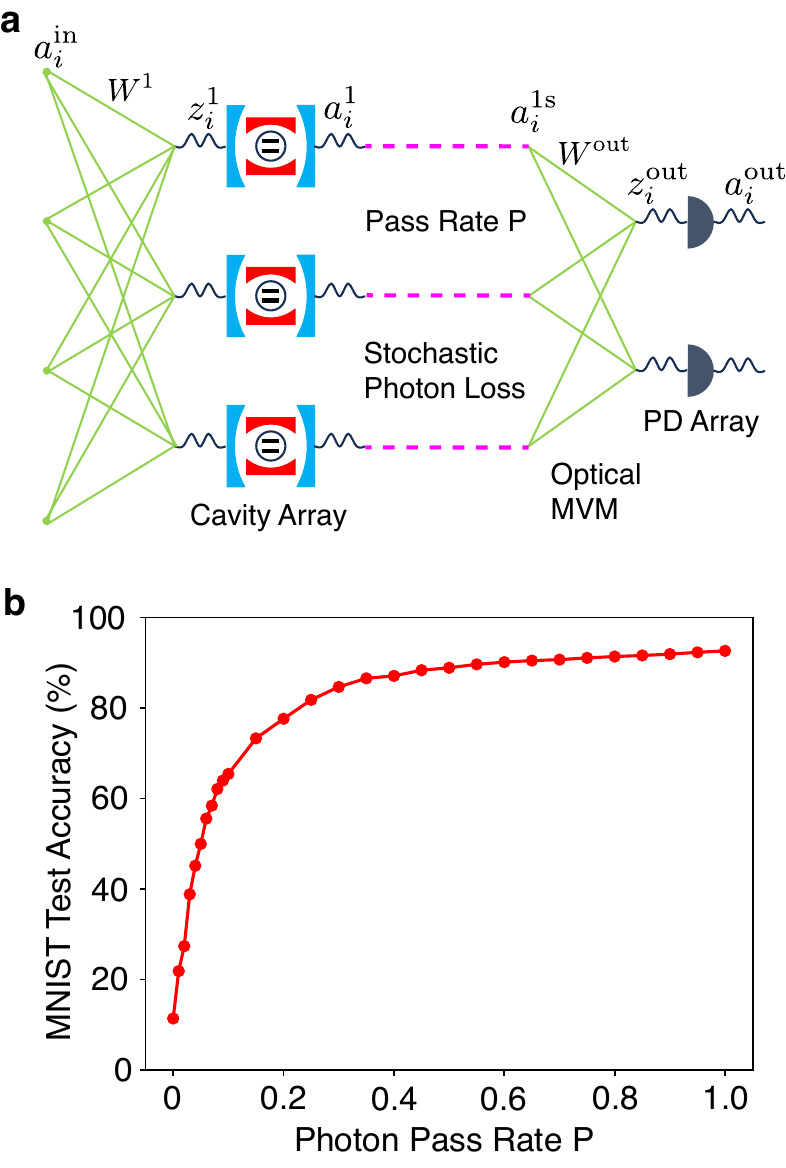}
\caption{Quantum optical neural network with stochastic photon loss. 
{\bf a} Based on the schematic shown in Fig.~\ref{Fig1}a, a stochastic layer (in magenta) is added before the activation $a_{i}^{1}$ entering the optical MVM to model photon loss during transmission. During the forward pass, each $a_{i}^{1}$ has a probability $P$ of remaining unchanged and a probability $1-P$ of being set to $0$. The photon passing rate $P$ is identical for all cavity neurons. 
{\bf b} MNIST test accuracy plotted against the photon passing rate.
The stochastic neural network consists of a single hidden layer with $512$ cavity neurons, each having a photon absorption time $t_{1} = 1$ and a detuning $\delta_{i}^{1} = 0$, and is trained for $10$ epochs.}
\label{Fig4}
\end{figure}

\subsection{Stochastic Photon Loss}

Stochastic photon loss is an inherent and practically unavoidable phenomenon in optical transmission systems \cite{Fedrizzi2009, Bonneau2015}. Figure~\ref{Fig4}a illustrates a stochastic QONN model to simulate the photon loss, where a stochastic layer is inserted before the emitted photons enter the optical MVM stage of the next layer. To simulate the single-photon loss semi-classically, we assume that each photon has a probability $P$ of passing through the stochastic layer unperturbed and a probability $1-P$ of being completely lost to the environment. This stochastic loss occurs independently for the photons emitted from different cavity neurons, with an identical transmission probability $P$ for all photons. During both the training and testing stages, the stochastic layer is included in the forward propagation. In the backward propagation during training, however, a mean-field approximation is employed, treating photon loss deterministically, as detailed in Appendix \ref{sec:Stochasticlayer}. The impact of stochastic photon loss on the MNIST classification task is presented in Fig.~\ref{Fig4}b, where a single hidden layer is considered for simplicity. Remarkably, the test accuracy remains around $80\%$ even when the photon pass rate decreases to $20\%$, demonstrating the strong robustness of the QONN against photon loss when performing classification tasks.

\begin{figure*}
\includegraphics[width=1.0\linewidth]{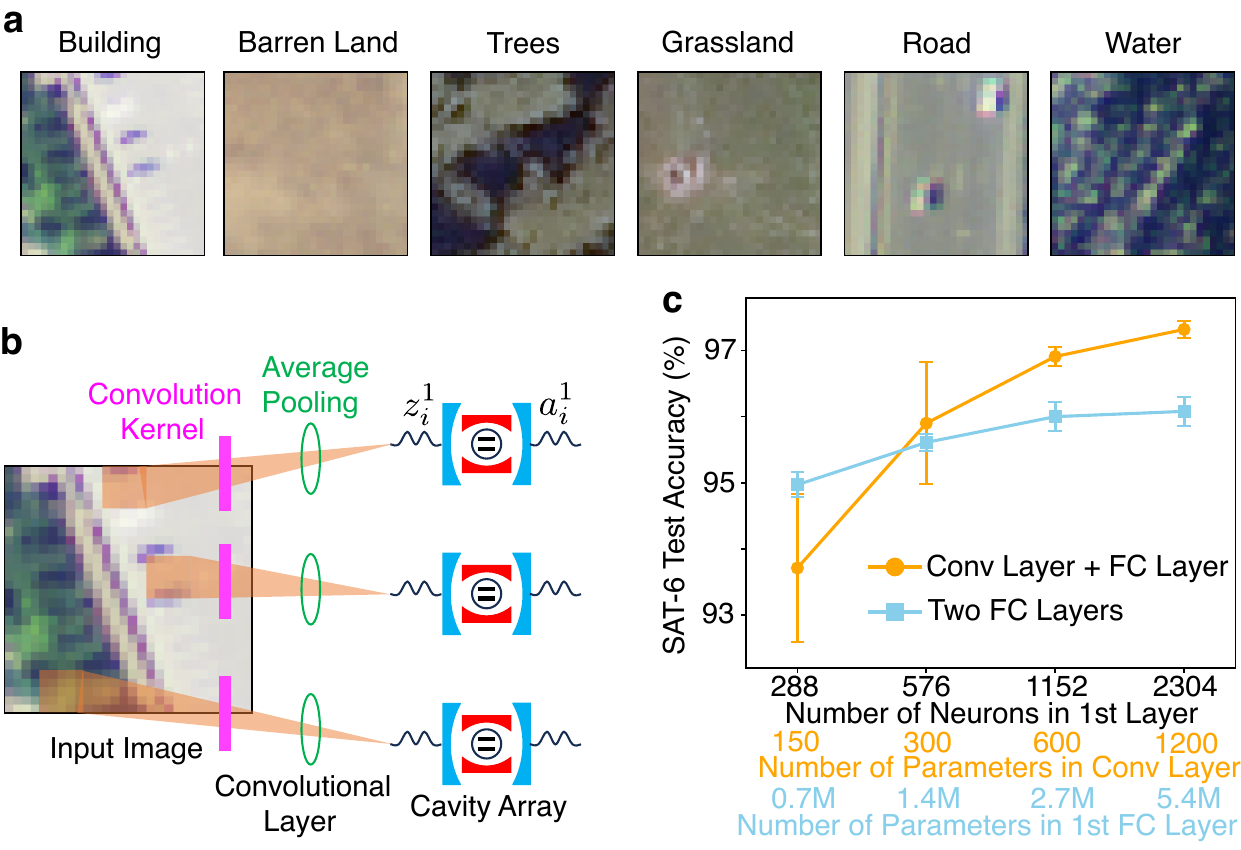}
\caption{Quantum optical neural network performing DeepSat (SAT-6) airborne image classification task. 
{\bf a} SAT-6 images with RGB channels representing six land cover classes. 
{\bf b} Schematic of a convolutional layer embedded in the quantum optical neural network, where the convolution kernel is implemented using a set of SLMs, followed by average pooling realized through programmable photonic circuits. The photons are incident into the cavity neurons after the convolution and pooling operations. 
{\bf c} Test accuracy of the quantum optical neural network on the SAT-6 task with varying numbers of cavity neurons in the first hidden layer. 
The error bars in {\bf c} represents the standard deviations across 5 independent runs initialized with different random seeds.
Two network structures are applied: one with a convolutional layer followed by a fully-connected layer (orange), and another with two fully-connected layers as shown in Fig.~\ref{Fig1}a (blue). The number of parameters in the convolutional layer depends on the kernel size and the number of output channels. A $5 \times 5$ kernel with a stride of $1$ and $2 \times 2$ average pooling with a stride of $2$ are used. The number of output channels in the convolutional layer is set to $2$, $4$, $8$, and $16$, corresponding to the four orange dots in panel {\bf c}, respectively. The second layer contains $512$ neurons in both structures. The network is trained for $10$ epochs.}
\label{Fig5}
\end{figure*}

\subsection{SAT-6 Task and Convolutional Neural Network}

The DeepSat (SAT-6) airborne image classification task \cite{Basu2015} is employed to demonstrate a real-world application of the QONN. The task is introduced in Appendix \ref{sec:Tasks}. Owing to its small hardware size and low-energy consumption, the QONN can serve as an on-board learning machine for real-time remote sensing on satellites. It reduces the communication bandwidth required between the satellite and ground stations, thereby enhancing both operational efficiency and data security. The task involves classifying satellite images into six land-cover categories, as illustrated in Fig.~\ref{Fig5}a.

To efficiently process images with RGB color channels, an optical convolutional layer is introduced, as shown in Fig.~\ref{Fig5}b, to reduce the number of trainable parameters. Unlike a fully connected (FC) layer, where all emitted photons are focused onto a single cavity, in the convolutional layer each cavity processes photons originating from a small spatial region (e.g., a $5\times5$ kernel in Fig.~\ref{Fig5}b). The same sets of convolutional filters are applied across multiple regions to extract shared features. The convolutional filter is performed by optical MVM with multiple SLMs, with the adjustable transmission rate of each pixel on SLMs treated as a parameter in the convolutional QONN. Three SLMs are encompassed in each set of filters to address the three color channels, and the number of the filter sets corresponds to the number of output channels. The average pooling, implemented by programmable photonic circuits \cite{Bogaerts2020}, combines the photon amplitudes after the convolution operation. After the convolution and pooling, photons are passed to the cavity array for nonlinear activations. 

Figure~\ref{Fig5}c plots the SAT-6 test result obtained using a convolutional QONN (in orange), which adopts the two-layer structure shown in Fig.~\ref{Fig1}a but replaces the first hidden layer with the convolutional layer depicted in Fig.~\ref{Fig5}b. For comparison, the original structure with two FC layers is also used to perform SAT-6 task using the same number of neurons (in blue). In the convolutional network, the number of parameters is determined by the kernel size and the number of output channels, with the details discussed in Appendix \ref{sec:Convlayer}. The FC network, however, requires a significantly larger number of parameters, which is decided by the product of the neuron counts in the two connected layers. The results in Fig.~\ref{Fig5}c highlight the advantage of the convolutional QONN, which substantially reduces the control complexity of SLM pixel transmission rates while maintaining comparable test accuracy without the need for additional cavity neurons.

\section{Discussion}

We have proposed a QONN to address the electronic delay and energy consumption issues in ONNs without the need for complex qubit gate design as seen in QNNs. The QONN harnesses quantum optical neurons to produce nonlinear activations through absorbing and emitting single photons, supporting both fully-connected and convolutional layers. The effects of photon absorption duration, random atom-photon detuning, and stochastic photon loss have been investigated. The QONN has achieved over $98\%$ accuracy on the benchmark MNIST digit classification task and over $97\%$ accuracy on the real-world SAT-6 task. This result surpasses the classical baselines with the typical accuracy $95\%$ for the MNIST task and $90\%$ for the SAT-6 task \cite{Basu2015}, and it is also comparable with the result from classical ONN experiments on the MNIST task with the accuracy roughly between $90\%$ and $98\%$ \cite{Ma2025}.

The expansion of the computational Hilbert space has been shown to enhance the capabilities of several quantum artificial intelligence frameworks, including quantum reservoir computing \cite{Zhu2025, Zhu2024}. One straightforward approach to enlarging the Hilbert space in a QONN is to incorporate photon-photon entanglement generated within the spatial light modulator or Mach-Zehnder interferometer through the inclusion of photon Fock states \cite{Steinbrecher2019, Ewaniuk2023, Roncallo2025, Killoran2019}. Such entanglement is not considered in our current semiclassical treatment of the incident and activation photon fields, as an analytical expression for the activation function, given by Eq.~(\ref{eq:activation}), is required for gradient-based training. Alternatively, within the present mean-field framework, a larger computational Hilbert space could be realized by introducing effective atom-atom entanglement among cavity neurons within the same network layer. This may be achieved by coupling the cavity neurons in the same layer via optical waveguides or by incorporating multiple atoms into each cavity. These directions warrant further investigation and may enable the discovery of more sophisticated, and potentially more efficient, quantum activation functions that achieve an improved balance between nonlinearity and monotonicity.

Due to its compact hardware and the potentials to reduce energy consumption and latency, the proposed QONN is suitable as an onboard learning system for real-time remote sensing on satellites, reducing the need for high-bandwidth communication with ground stations and thereby improving data security.
The proposed QONN can also find potential applications in optoelectronic devices, enabling high-speed and energy-efficient information processing through photonic hardware. It can potentially be used to enhance optical communication systems by improving signal processing and noise mitigation, support machine-learning tasks such as pattern recognition and classification, and improve the performance of optoelectronic sensing and imaging systems.


\section*{Acknowledgements}
This work is supported by the U.S. Department of Energy, Office of Science, Office of Biological and Environmental Research under Award Number DE-SC0025910.

\appendix

\section{MNIST and SAT-6 Tasks}
\label{sec:Tasks}

The benchmark MNIST (Modified National Institute of Standards and Technology) handwritten digit dataset comprises $60000$ training images and $10000$ test images \cite{Deng2012}. Each image is a $28\times28$ grayscale matrix, resulting in $784$ pixels per image. To encode the image onto the input photon amplitude, all pixel values are normalized to fall within the range $\left[0,1\right]$.

The real-world DeepSat (SAT-6) airborne image classification task consists of $324000$ training images and $81000$ test images, which fall into six land cover classes including buildings, barren land, trees, grassland, roads, and water bodies \cite{Basu2015}. Each image consists of $28\times28$ pixels and $4$ color channels - red, green, blue (RGB) and near infrared. We only use the RGB channels, leading to $2352$ data points per image. 

\section{Backward Propagation for Training}
\label{sec:Backpropagation}

The cost function $C\left(a_{i}^{{\rm out}}\right)$ characterizes the error of the inference by comparing the output $a_{i}^{{\rm out}}$ with the ground truth. We employ the cross-entropy loss method \cite{Mao2023}, which applies LogSoftmax function on $z_{i}^{{\rm out}}$ to simulate the nonlinearity induced by the photodetector array in the output layer, followed by the negative log likelihood loss (NLLLoss) \cite{Yao2020} to evaluate the cost function. The backward propagation is processed with chain rule as
\begin{align}
\frac{\partial C}{\partial z_{i}^{l}} & =\frac{\partial a_{i}^{l}}{\partial z_{i}^{l}}\frac{\partial C}{\partial a_{i}^{l}},\label{eq:dCdz}\\
\frac{\partial C}{\partial a_{i}^{l-1}} & =\underset{j}{\sum}\frac{\partial C}{\partial z_{j}^{l}}W_{ji}^{l},\label{eq:dCda}
\end{align}
where $\partial a_{i}^{l}/\partial z_{i}^{l}$ is analytically expressed as the derivative of the activation function $a_{i}^{l}\left(z_{i}^{l}\right)$ given by Eq.~(\ref{eq:activation}). For example, the layers in Fig.~\ref{Fig1}a are processed backward in the order of $l={\rm out},2,1,{\rm in}$. The gradient of the cost function with respect to the weight $W_{ij}^{l}$ is computed as 
\begin{equation}
g_{W_{ij}^{l}}\equiv\frac{\partial C}{\partial W_{ij}^{l}}=\frac{\partial C}{\partial z_{i}^{l}}a_{j}^{l-1},\label{eq:dCdW}
\end{equation}
where $\partial C/\partial z_{i}^{l}$ is given by Eq.~(\ref{eq:dCdz}). After running through a batch of training images, we use stochastic gradient descent (SGD) optimizer to update the weights based on the gradient values obtained from Eq.~(\ref{eq:dCdW}). During the training, we restrict the value of $W_{ij}^{l}$ in the range $[-1,1]$ on account of the actual photon transmission rate of a SLM pixel. We set the batch size to $64$ and the learning rate to $10^{-3}$, and use PyTorch in the simulation.

\section{Stochastic Layer}
\label{sec:Stochasticlayer}

The photon pass rate $P$ is introduced to characterize the single-photon loss in the stochastic layer. As shown in Fig.~\ref{Fig4}a, in the forward propagation for training and testing, the stochastic photon loss is simulated as 
\begin{equation}
a_{i}^{{\rm 1s}}=f\left(P\right)a_{i}^{{\rm 1}},\label{eq:a1sforward}
\end{equation}
where $f\left(P\right)$ is a random variable with a Bernoulli distribution: $f\left(P\right)=1$ with probability $P$, and $f\left(P\right)=0$ with probability $1-P$. In the backward propagation for training, the deterministic mean-field approximation is implemented by assuming
\begin{equation}
a_{i}^{{\rm 1s}}=Pa_{i}^{{\rm 1}}.\label{eq:a1sbackward}
\end{equation}
This correspondingly modifies the gradient in the stochastic QONN as
\begin{equation}
g_{W_{ij}^{l}}^{{\rm stochastic}}=Pg_{W_{ij}^{l}},\label{eq:gstoc}
\end{equation}
where $g_{W_{ij}^{l}}$ is defined in Eq.~(\ref{eq:dCdW}). The modified gradient is involved in the weights update through the SGD optimizer for both layers indexed by $l={\rm out}$ and $l=1$ in Fig.~\ref{Fig4}a. 

The proposed stochastic photon loss model can be viewed as a beam-splitter loss model. In the beam-splitter description, a mode with photon loss undergoes the transformation $a_{i}^{1{\rm s}}=\sqrt{p}a_{i}^{1}+\sqrt{\left(1-p\right)}v$, where $v$ is a vacuum mode. This channel corresponds to the independent survival of the photon with probability $p$, which is related to the photon passing rate $P$ by $P=\sqrt{p}$. Rather than propagating the full quantum state, we employ a mean-field approximation in which a binary variable is sampled as $f(P)$. During backpropagation, the stochastic mask is replaced by its expectation, giving $f(P)=P$. This provides a semiclassical approximation to the expected gradient while retaining the correct average effect of photon loss.

\section{Convolutional Layer}
\label{sec:Convlayer}

In the convolutional layer illustrated in Fig.~\ref{Fig5}b, the input SAT-6 image offers a feature map of $28\times28\times3$. It is followed by a convolution operation with a kernel size of $5\times5$, a stride size of $1$, a padding of $0$, and $N_{{\rm channels}}$ output channels, which results in a data dimension of $24\times24\times N_{{\rm channels}}$. After that, an average pooling of $2\times2$ is applied to reduce the data dimension to $12\times12\times N_{{\rm channels}}$, which is then flattened into a vector and applied to the neurons in the first layer of the network. In Fig.~\ref{Fig5}c, we consider $N_{{\rm channels}}=2,4,8,16$, which necessitates the number of neurons from $288$ to $2304$. The number of parameters in the convolutional layer is given by $5\times5\times3\times N_{{\rm channels}}$ with the $5\times5$ kernel and $3$ input color channels. This number varies from $150$ to $1200$ in Fig.~\ref{Fig5}c.


\bibliography{References_QONN}

\end{document}